\documentclass[
 reprint,amsmath,amssymb,aps,prb,eps,superscriptaddress,twocolumn,10pt
]{revtex4-2}

\usepackage{graphicx}
\usepackage{dcolumn}
\usepackage{grffile}
\usepackage{bm}
\usepackage[colorlinks=true,
            linkcolor=blue,
            urlcolor=blue,
            citecolor=blue]{hyperref}
\usepackage{xcolor}
\usepackage{placeins}
\usepackage{changes}
\usepackage{enumitem}
\usepackage{blindtext}

\definechangesauthor[name=Markus Aichhorn, color=red]{MA}
\definechangesauthor[name=Johannes Graspeuntner, color=blue]{JG}

\begin{document}

\preprint{APS/123-QED}

\title{Non-local self energies in pyrochlore iridates from \textit{ab-initio} TRILEX calculations, and their relevance for the Weyl semimetal phase}

\author{Johannes Graspeuntner}
\affiliation{Institute of Theoretical and Computational Physics, Graz University of Technology, NAWI Graz, Petersga{\ss}e 16, Graz, 8010, Austria.}
\author{Markus Richter}
\affiliation{Institute of Theoretical and Computational Physics, Graz University of Technology, NAWI Graz, Petersga{\ss}e 16, Graz, 8010, Austria.}
\author{Markus Aichhorn}
\email{aichhorn@tugraz.at}
\affiliation{Institute of Theoretical and Computational Physics, Graz University of Technology, NAWI Graz, Petersga{\ss}e 16, Graz, 8010, Austria.}

\newcommand{\appropto}{\mathrel{\vcenter{
  \offinterlineskip\halign{\hfil$##$\cr
    \propto\cr\noalign{\kern2pt}\sim\cr\noalign{\kern-2pt}}}}}

\date{\today}

\begin{abstract}
Motivated by recent experiments and computational results on pyrochlore iridates, we compare single-particle properties of Y$_2$Ir$_2$O$_7$ obtained from 
single-site dynamical mean-field calculations with results within the TRILEX approximation, where the latter takes non-local correlations into account. Our calculations are all based on ab-initio calculations within density-functional theory, and take spin-orbit coupling into account. In order to make the treatment within TRILEX feasible, we first define a single-band $j_\textrm{eff}=1/2$ model, by comparing its spectral features within DMFT to a three-band model that includes both $j_\textrm{eff}=1/2$ and $j_\textrm{eff}=3/2$ orbitals. Our calculations show consistently a paramagnetic metallic phase at small interaction values, and an 
insulating antiferromagnetic phase at larger interaction values. The critical interactions, however, differ between single-site and TRILEX calculations. The antiferromagnetic phase shows the already predicted all-in/all-out magnetic ordering. Different to the single-site results, the TRILEX calculation gives also evidence for the Weyl-semimetal regime in the vicinity of the metal-insulator transition. 
\end{abstract}

\maketitle

\section{Introduction}

In recent years, a lot of effort has been devoted to the investigation and understanding of spin-orbit coupling (SOC) and its interplay with strong electronic correlations~\cite{krempa_review}. Among those effects that rise from SOC in strongly correlated systems, are SOC-driven Mott insulators~\cite{moon,martins_2011,martins_2017,zhang_2013}, new phases such as chiral spin liquids~\cite{chiral_spin_liquids,nature_chiral_spin_liquid}, axion insulators~\cite{WSM_and_AI_in_PyrochloreIridates,Topo_and_mag_in_PyrochloreIdridates,PyrochloreIridates_under_pressure} and Weyl semi-metals (WSM)~\cite{WSM_and_AI_in_PyrochloreIridates,Corr_effects_in_3D_topo_phase}.  In particular, pyrochlore-iridates R$_2$Ir$_2$O$_7$ (where R represents rare-earth elements or Yttrium) have been studied very closely in this respect. In these systems, similar to other iridate compounds, the presence of SOC splits the t$_{2g}$ orbitals into $j_\textrm{eff}=3/2$ and $j_\textrm{eff} = 1/2$ orbitals. With the $j_\textrm{eff}=1/2$ orbital being close to half-filled, these systems show particularly strong effects of electronic correlations, leading to a strongly-correlated spin-orbital polarized state~\cite{martins_2011} that eventually leads to insulating behavior in quite a number of iridate compounds.

First calculations for Y$_2$Ir$_2$O$_7$ within density-functional theory plus mean-field-like correlations (DFT+U) predicted a WSM with an all-in/all-out (AIAO) magnetic ordering~\cite{WSM_and_AI_in_PyrochloreIridates}. Later on, Hartree-Fock-based calculations that use a tight-binding model description~\cite{Topo_and_mag_in_PyrochloreIdridates,Topo_and_mag_in_PyrochloreIdridates}, as well as results from cluster dynamical mean-field (CDMFT) theory~\cite{Corr_effects_in_3D_topo_phase,CDMFT_weyl_rings_in_PyrochloreIridates} were concluding with a similar picture for the existence of a WSM. The region in parameter space, however, was very small~\cite{CDMFT_weyl_rings_in_PyrochloreIridates}.
In contrast to those works, the combination of DFT with single site dynamical mean-field (sDMFT) has lead to different conclusions in one important aspect. In these calculations~\cite{sDMFT_soc_and_corr_in_PyrochloreIridates,PhysRevLett.118.026404,DFT+DMFT_Werner}, the system Y$_2$Ir$_2$O$_7$ undergoes a direct transition from a paramagnetic metal to a topologically trivial AIAO antiferromagnetic insulator, without the appearance of a WSM-phase in between.

Experimental works have studied the pyrochlore iridates in dependence of the rare earth element. By exchanging the rare-earth ion along the rare-earth series, a metal-insulator transition has been found~\cite{Metal_Nonmetal_transition_exp,Metal_Insulator_Ln2Ir2O7_exp,Magnetic_transition_Eu2Ir2O7_exp,Cont_transition_Eu2Ir2O7_exp,Mag_order_Y2Ir2O7_exp,Mag_Nd2_Ir2_O7_exp,optical_conductiviy_metal_insulator_pyrochlore_exp}.  At the same time, the system exhibits a phase transition from a paramagnetic to an antiferromagnetic state with AIAO ordering, where the iridium atoms are occupying a frustrated lattice, and time-reversal symmetry is broken. From that, a WSM-phase is likely to exist and has shown up experimentally in optical response measurements~\cite{WSM_EU2Ir2O2_exp}. Also the measurement of the spontaneous Hall effect in pyrochlore iridates under the application of an external magnetic field confirms the existence of a WSM phase~\cite{H_induced_WSM_exp,Spont_Hall_in_WSM_exp}. 
However, measurements by angle-resolved photoemission spectroscopy (ARPES) were, not able to detect Weyl points~\cite{Slater_Mott_Crossover_Nd2Ir2O7_exp}. 

Coming back to theory, a comparison of the results of  sDMFT~\cite{sDMFT_soc_and_corr_in_PyrochloreIridates,PhysRevLett.118.026404,DFT+DMFT_Werner} and CDMFT~\cite{CDMFT_weyl_rings_in_PyrochloreIridates} calculations suggests that the WSM phase arises due to non-local self-energies, which are neglected in sDMFT calculation by construction. In this paper, we take a closer look at the momentum dependence of the self-energy in Y$_2$Ir$_2$O$_7$. We want to investigate the possible existence of a WSM phase near the metal-insulator transition within the framework of density functional theory combined with the triply-irreducible leg extension (DFT+TRILEX). In contrast to sDMFT, the TRILEX method \cite{TRILEX,TRILEX_2}
does not approximate the self energies as local quantities, but the next higher-order diagram, namely the triple legged $\Lambda$ diagram that connects the fermionic and bosonic states. We will show that there is sizeable momentum dependence in the self-energy, which makes the band structure and also the phase diagram different from sDMFT results.

\section{Methods and Low-Energy Modelling}

\begin{figure}
	\begin{center}
		\includegraphics[width=0.7\columnwidth]{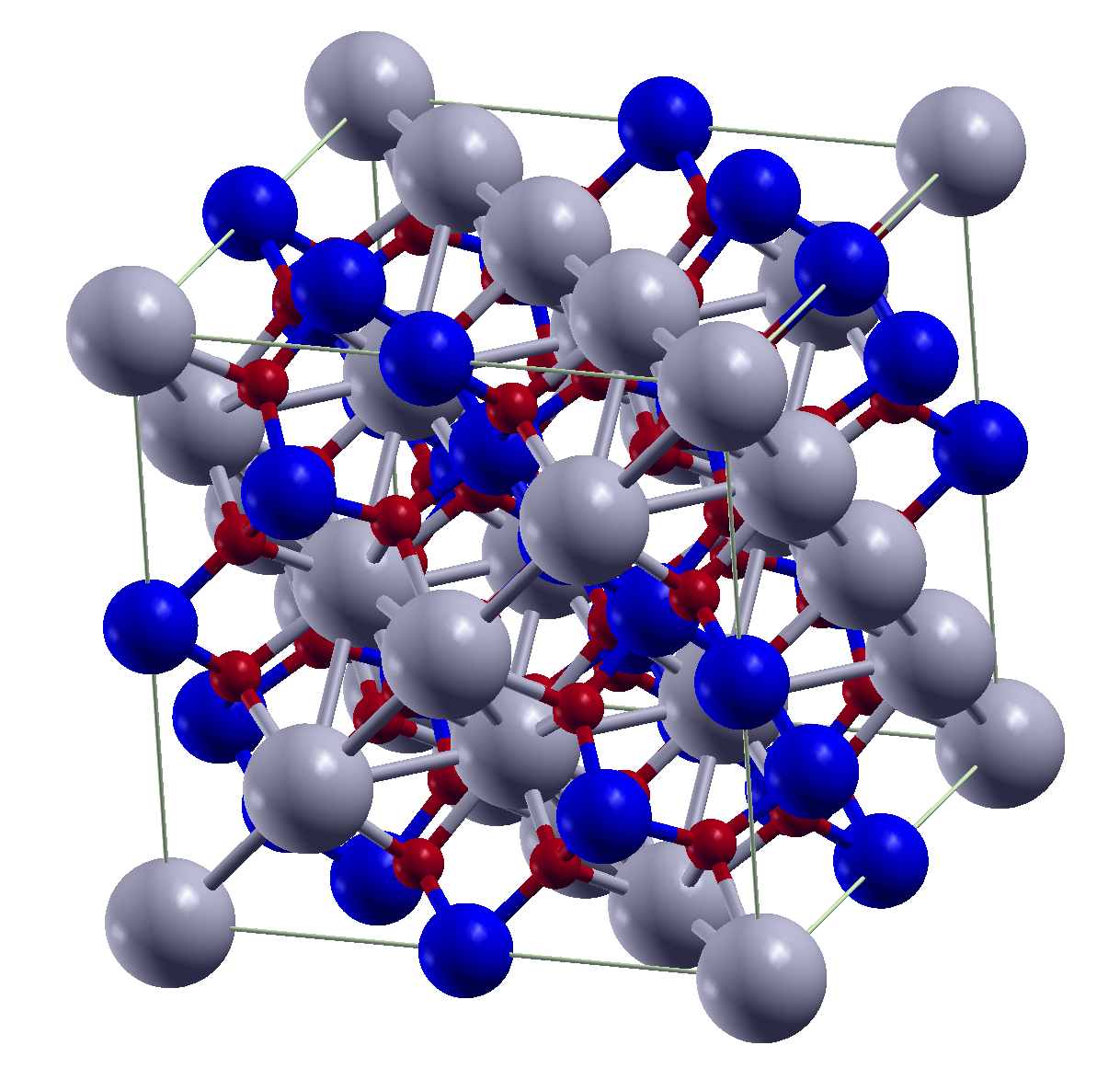}
		\vspace*{0.5cm}
		\includegraphics[width=0.8\columnwidth]{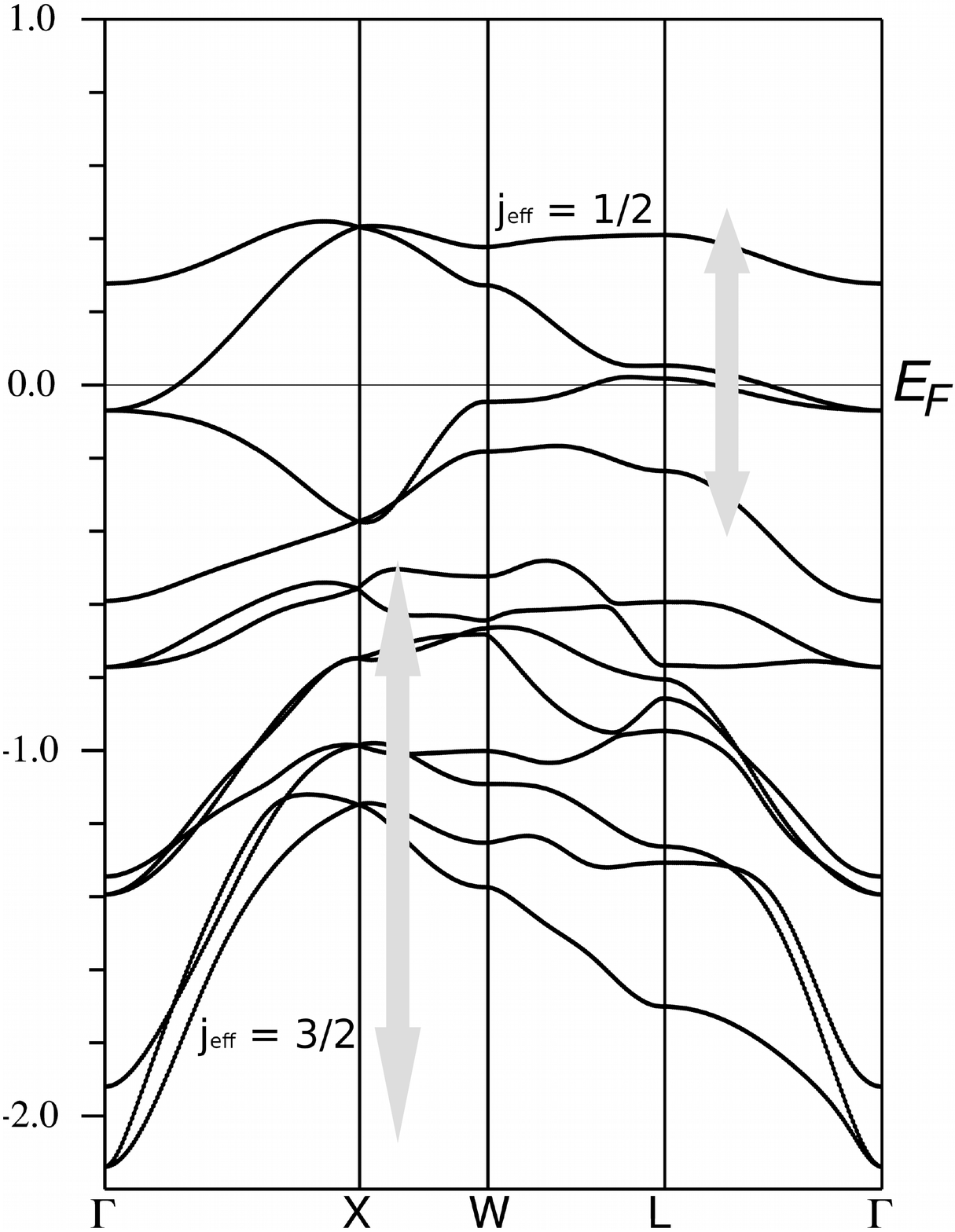}
		\caption{Top: Crystal structure of the pyrochlore iridate, Y$_2$Ir$_2$O$_7$. The large blue spheres depict the iridium atoms. The large grey spheres stand for the rare earth elements; Yttrium in this case. The small red spheres are the oxygen atoms. 
		Bottom: The bandstructure of Y$_2$Ir$_2$O$_7$ obtained from DFT+soc calculations, where only the energy range of the iridium t$_{2g}$ orbitals is shown. It can be clearly seen that the the bands split into a $j_\textrm{eff}=1/2$ and a $j_\textrm{eff}=3/2$ manifold in the presence of SOC.  }
		\label{fig:crystal_struct_band_struct}
	\end{center}
\end{figure}

In Fig.~\ref{fig:crystal_struct_band_struct} we show the crystal structure of the pyrochlore-iridate family. 
Generally speaking, pyrochlore iridates are described by the chemical formula R$_2$Ir$_2$O$_7$, where the R stands for the rare-earth elements. In this paper we are fixing this element to Yttrium.  
The Ir atoms are arranged in a corner sharing tetrahedral structure, forming a frustrated lattice. Close to the Fermi energy only the electrons of the Ir 5d shell are present. Hence, we have five valence electrons in the t$_{2g}$ orbitals that are affected by strong on-site correlations as well as strong spin-orbit coupling.

To describe the physics of the system we perform DFT+sDMFT \cite{triqs} as well as DFT+TRILEX \cite{TRILEX,TRILEX_2} calculations. The first step is to do the ab-initio calculation, where we use the Wien2k package~\cite{wien2k}. As exchange-correlation functional the generalized gradient approximation in the parametrization of the Perdew, Burke, and Ernzerhof (PBE) has been used. In the calculations we are using a $8\times 8\times 8$ k-point grid. Due to the presence of spin-orbit coupling in this strongly correlated system the t$_{2g}$ orbitals are further split into a $j_\textrm{eff} = \frac{1}{2}$ and a $j_\textrm{eff} = \frac{3}{2}$ manifold, where the latter is filled up to 90$\%$, while the $j_\textrm{eff} = \frac{1}{2}$ manifold takes the remaining charge-density and shows an occupancy of $1.4$  electrons. Thus, the spin-orbital polarisation in the non-interacting case is not complete, similar to other iridate compounds~\cite{martins_2011}.

The bandstructure plot of the DFT calculation in Fig.~\ref{fig:crystal_struct_band_struct} depicts a clear separation of the $j_\textrm{eff} = \frac{1}{2}$ and the $j_\textrm{eff} = \frac{3}{2}$ manifold in reciprocal space. This separation suggests that it should also be possible to approximate the system by a half-filled one-band model. In fact, such an approximation is essential in order to reduce the complexity of the problem and the computational effort for a treatment of many-body correlations in the TRILEX approach, which we will use in this paper. In the close vicinity of the $j_\textrm{eff} = \frac{3}{2}$ manifold the physics of the $j_\textrm{eff} = \frac{1}{2}$ manifold is definitely affected. To correctly treat the low energy physics, the system is projected on maximally localized wannier functions (MLWF) with the wannier90 code \cite{wannier90}.  

The Wannier-projected Hamiltonian is used to perform single-site DMFT calculations for both the three band t$_{2g}$ model as well as the $j_\textrm{eff} = \frac{1}{2}$ model, in order to compare the qualitative behaviour. The calculations of the t$_{2g}$ model have been performed using a Kanamori Hamiltonian with the Hund's coupling being $J=0.1U$. For the one band case of the the $j_\textrm{eff} = \frac{1}{2}$ manifold, a Hubbard-type interaction Hamiltonian with parameter $U$ is sufficient. All calculations are done using the TRIQS toolkit~\cite{triqs}. For the sDMFT calculations we use the TRIQS/CTHYB solver~\cite{CTHYB_1,CTHYB_2,CTHYB_3}. 

\begin{figure}
	\begin{center}
		\includegraphics[width=1.0\columnwidth]{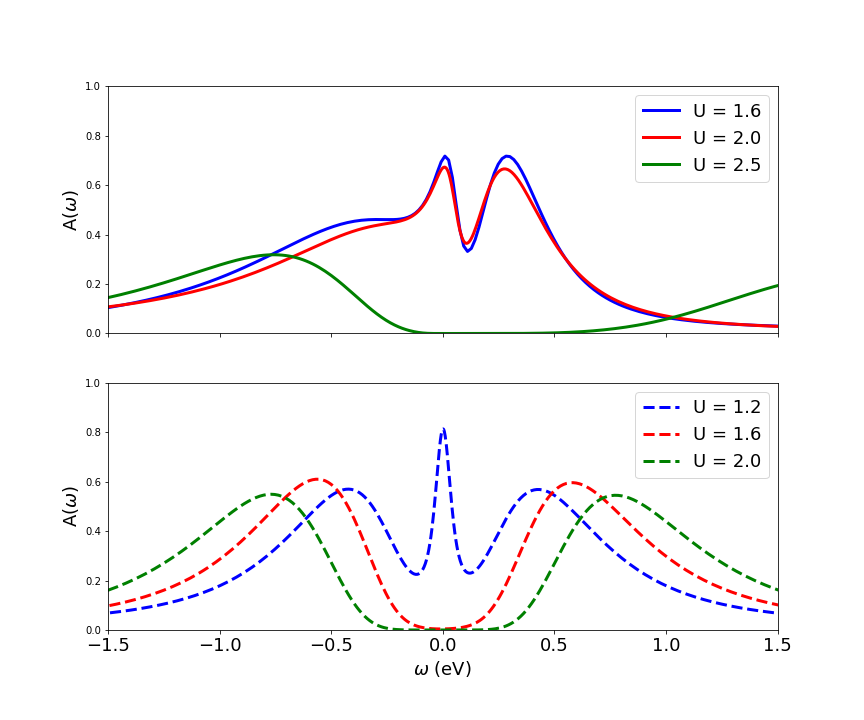}
		\caption{Top: Spectral function of the $j_\textrm{eff}=\frac{1}{2}$ orbital in the full t$_{2g}$-band calculation and for three different interaction values $U=\{ 1.6, 2.0,2.5 \}$\,eV and $J = 0.1U$. The asymmetry up to $U=2.5$\,eV stems from the still incomplete spin-orbital polarization in that case. This polarization becomes complete for larger interactions, hence, the spectra become more symmetric.
		Bottom: The spectral function for the $j_\textrm{eff}=\frac{1}{2}$ one-band model and three interaction values. All calculations are done at $\beta=40$\,eV$^{-1}$, and spectra are obtained using the TRIQS/MaxEnt package~\cite{maxent}.
  }
		\label{fig:A_t2g_j12}
	\end{center}
\end{figure}

The comparison of the models was performed at higher temperatures, $\beta = 40$\,eV$^{-1}$, in order to avoid possible sign problems of the multi-band calculations, that could arise with the CTHYB solver. Performing multi band calculations over a range of different interaction values $U$ (with $J=0.1U$) show that the metal-insulator Mott-transition occurs at an interaction value close to $U \approx 2.5$\,eV , see Fig.~\ref{fig:A_t2g_j12}, which is in line with previous treatments in literature~\cite{DFT+DMFT_Werner}. This transition is accompanied by a complete spin-orbital polarization with filled $j_\textrm{eff}=\frac{3}{2}$ and half-filled $j_\textrm{eff}=\frac{1}{2}$ orbitals. What the comparison of the multi-band and the one-band model shows is that the metal insulator transition of the $j_\textrm{eff} =\frac{1}{2}$ model occurs at a reasonable value as well.

To treat the longer-range correlations we are using the triply-irreducible local expansion (TRILEX) framework \cite{TRILEX,TRILEX_2}. Same as for the sDMFT calculation we use the Wannier projected Hamiltonian and add on-site interactions to the Ir atoms. In order to interface the Wannier projection with the TRILEX calculation, we use again the TRIQS package~\cite{triqs}, in particular the lattice tools application. 
In comparison to the DMFT approach, TRILEX treats the self-energy k-dependent and makes the approximation with a local three-index-vertex $\Lambda_{imp}(i\omega,i\Omega)$.
Hence the TRILEX method interpolates between the fluctuation exchange limit, that treats long-range fluctuations, and the DMFT limit, that approximates the self-energy as a local frequency dependent impurity self-energy. For the calculation we have chosen to use $\alpha = 0.5$ as the decoupling parameter between the spin and the charge channels of the Hubbard interaction, based on the fact that the different alpha values had very little influence on the result. 
Since we deal with a symmetry-broken antiferromagnetic system, we use the specific approximation for antiferromagnetic cases called the $\Lambda ^2$-TRILEX derived in ~\cite{Richter}. For the TRILEX calculations we are using the CTINT solver~\cite{CTINT}. 
This had the advantage of the linear error in the tails of the Green's functions. The quadratic error of the CTHYB 
solver would lead to numerical errors in the Hedin equations. 

For the TRILEX calculations the computational efforts become infeasible near the phase transition, due to the critical slowing down of Monte Carlo processes near the phase transitions, which impacts the resolution of the bosonic green's functions drastically and as a result also impacts the other observables. If the Green's functions show to much noise, the calculations eventually fail when trying to perform Fourier transforms. As a result we were able to perform converged self-consistent TRILEX calculations up to an interaction value of $U=1.2$\,eV with low enough numerical noise to obtain useful results. For higher interaction values, in order to probe the k-dependence and the possible WSM phase, we limited the calculation to one-shot TRILEX calculations. For that, fully converged DMFT solutions have been used to calculate the vertex corrections just once in order to obtain insight into non-local physics. 

\section{Results}

\subsection{Magnetic Properties}

As stated in the previous section, all calculations that we present in the following are obtained from the effective $j_\textrm{eff}=\frac{1}{2}$ one-band model. We start with DMFT calculations at a certain interaction value $U = 2.0$\,eV in the insulating regime. We already know from the calculations presented in the previous section, see Fig.\ref{fig:A_t2g_j12}, 
that this value is well beyond the phase transition point. Starting out from a rather high temperature $\beta = 40$\,eV$^{-1}$, we perform a series of calculations at lower temperatures to determine an optimal temperature in where we can see a substantial magnetic ordering and at the same time keep the computational costs at a reasonable level. 

Based on these considerations we found that doing calculations at $\beta = 180$\,eV$^{-1}$ is a good trade off that works for different interaction values. As the interaction value increases, the system undergoes not only a metal-insulator transition, but also a paramagnetic to antiferromagnetic transition. 
It starts out from a paramagnetic metallic phase and progresses to a all-in/all-out ordered antiferromagnetic metal phase. By further increasing the interaction value it transforms to an antiferromagnetic insulator. This result is coherent with the single-site multi-orbital calculations~\cite{sDMFT_soc_and_corr_in_PyrochloreIridates,zhang_2013,DFT+DMFT_Werner}, where the system undergoes a direct Mott-transition from metal to insulator without any topological phase in between. In the plot of the bandstructure, which we will present later, we will see the gradually more compressed bands up to the critical interaction value, after which the correlations open the Mott gap in the insulting phase. 

In order to determine the critical interaction value where the magnetic transition occurs, we performed calculations for both the paramagnetic and the antiferromagnetic solutions and compared the respective ground state energy of the phases. We calculate the energy using the Migdal formula, 
\begin{equation}
\label{eq:Energy}
E =  \frac{1}{N_k\beta}\sum_{n}\sum_{k}\textrm{Tr}\{[{H}_0(k) + \frac{1}{2}\Sigma(i\omega_n)]{G}(k,i\omega_n)\}\, ,
\end{equation}
where $N_k$ is the number of k-points on the grid, $H_0(k)$ the non-interacting Wannier Hamiltonian, $\Sigma(i\omega_n)$ the local DMFT self energy, and $G(k,i\omega)$ the interacting DMFT Greens function.
\begin{figure}
	\begin{center}
		\includegraphics[width=0.9\columnwidth]{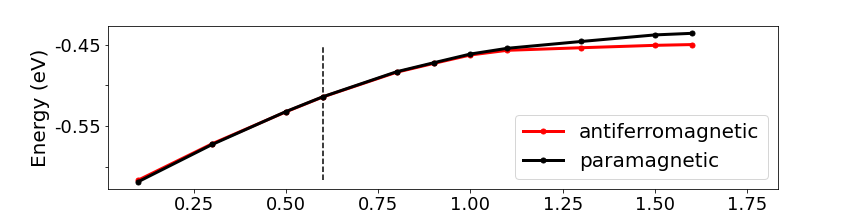}
		\vspace*{0.5cm}
		\includegraphics[width=0.9\columnwidth]{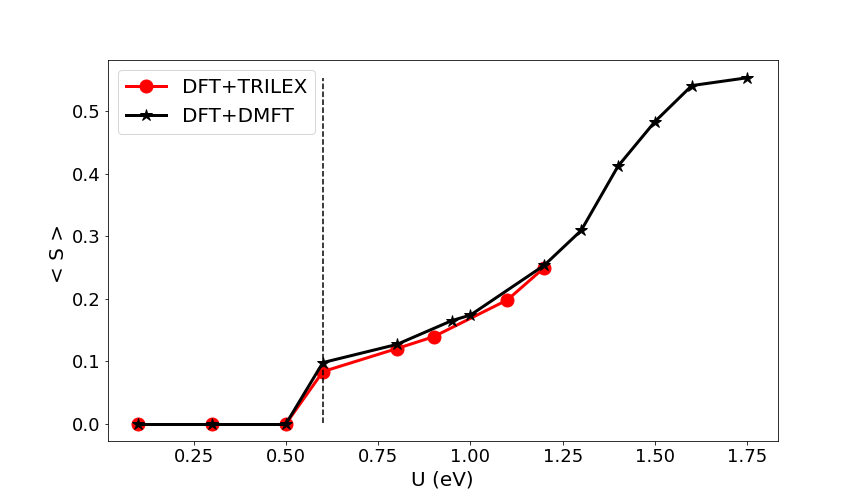}
		\caption{Top: The energy Eq.~\eqref{eq:Energy} plotted versus interaction strength for the sDMFT calculations at $\beta = 180$\,eV$^{-1}$. 
		The vertical line denotes the first-order phase transition at $U\approx 0.6$\,eV.
		Bottom: The expectation value of the spin operator $\langle S\rangle$ as function of the interaction strength, where the black line denotes the sDMFT result and the red line gives the results for the fully converged TRILEX calculations.}
		\label{fig:Energy_DMFT}
	\end{center}
\end{figure}

With this procedure we can determine the critical interaction for the magnetic phase transition to $U\approx 0.6$\,eV. 
Knowing where the ordered phase is starting it is reasonable to determine the expectation value of the spin operator $\langle \vec{S} \rangle$. The size of the magnetic moment is obtained via $\sqrt{\langle S_x \rangle^2 + \langle S_y \rangle^2 + \langle S_z \rangle^2}$, and we plot it against the interaction values $U$. Figure~\ref{fig:Energy_DMFT} shows a saturation of the moment for large interaction values.

The magnetisation plot of the TRILEX solution is very similar to the sDMFT one. As shown in Fig.~\ref{fig:Energy_DMFT}
they are basically aligning on top of each other. This result is expected, since other works have shown similar magnetic behaviour regardless whether k-dependent or independent methods were used.

\subsection{Momentum dependence of the self energy}

In this section we want to take a closer look at the k-dependence of the self energy of Y$_2$Ir$_2$O$_7$. 
The self energies from sDMFT calculations are obviously independent of momentum, since that is the approximation we make when using single-site DMFT. On the other hand, CDMFT takes some non-locality into account by using clusters of finite size. From the work of Wang and coworkers~\cite{CDMFT_weyl_rings_in_PyrochloreIridates} we see that the WSM phase occurs in CDMFT. Therefore, we can assume that the WSM phase can only be described by a momentum dependent self energy, and, thus, a method that is taking non-locality into account, is needed for a proper description.
The method we use here is the TRILEX approximation, which approximates a higher-order property as local, and therefore includes non-local effects terms in the self energy. 

\begin{figure}
	\begin{center}
		\includegraphics[width=0.9\columnwidth]{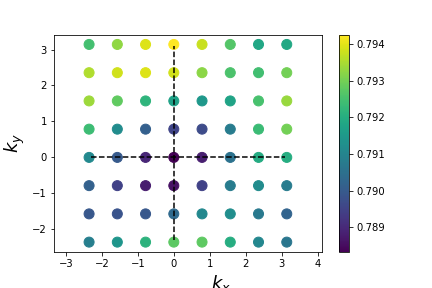}
		\caption{Momentum dependence of the real part of the self energy for $U=1.2$\,eV, at the first Matsubara frequency.}
		\label{fig:K_dependance_120}
	\end{center}
\end{figure}

For small interaction strength the k-dependence of the self-energy is negligible, similar to the sDMFT calculations. Up to a value of $U = 1.2$\,eV the self energy remains very local and we see little variance of its value over k-space. In Fig.~\ref{fig:K_dependance_120} we show the dependence of the real part of the self energy at the first Matsubara frequency in k space. The difference between the maximal and the minimal value is just of the order of $10^{-2}$\,eV.

\begin{figure}
	\begin{center}
		\includegraphics[width=0.9\columnwidth]{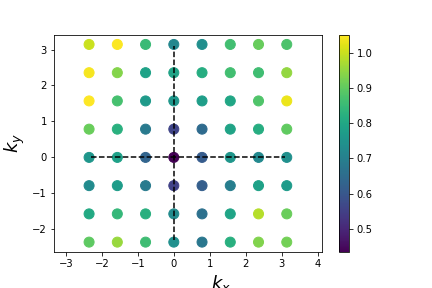}
		\caption{Momentum dependence of the real part of the self energy for $U=1.4$\,eV, at the first Matsubara frequency.}
		\label{fig:K_dependance_120_2}
	\end{center}
\end{figure}

We now increase the interaction strength in order to approach the phase transition point. However, for values of $U$  
where we would expect the WSM phase to be, the Monte Carlo steps needed in the CTINT solver to produce reasonable results with acceptable numerical noise increase dramatically. Facing such high computational costs, we decided to perform what we call one-shot TRILEX calculations. Instead of performing the full self-consistent TRILEX loops, which is still possible for values up to $U=1.2$\,eV, we now use a fully converged sDMFT solution as input to calculate the vertex $\Lambda$ just once, and investigate whether it already has an effect on the k dependence of the self energy. In addition, the symmetry around $k = [0,0,0]$ of the self energy is dependent on the smoothness of the polarisation function in the Hedin equations. Achieving good enough QMC quality to make the polarization smooth is also getting very costly near the phase transition.

We show the results for a single shot TRILEX calculation at $U=1.4$\,eV in Fig.~\ref{fig:K_dependance_120_2}, we can see a very strong variance of the self energy in momentum space. Compared to the k-dependence at $U=1.2$\,eV, where the self energy only varies at third decimal, the variations of the self energy at $U=1.4$\,eV are two orders of magnitude larger.

As stated above, the strong k-dependence in this interaction range is important for the presence of a WSM phase. To verify the presence of a WSM phase within the TRILEX framework we will take a look at the bandstructures in the next section. 

\subsection{Bandstructures}

\begin{figure}
	\begin{center}
		\includegraphics[width=0.9\columnwidth]{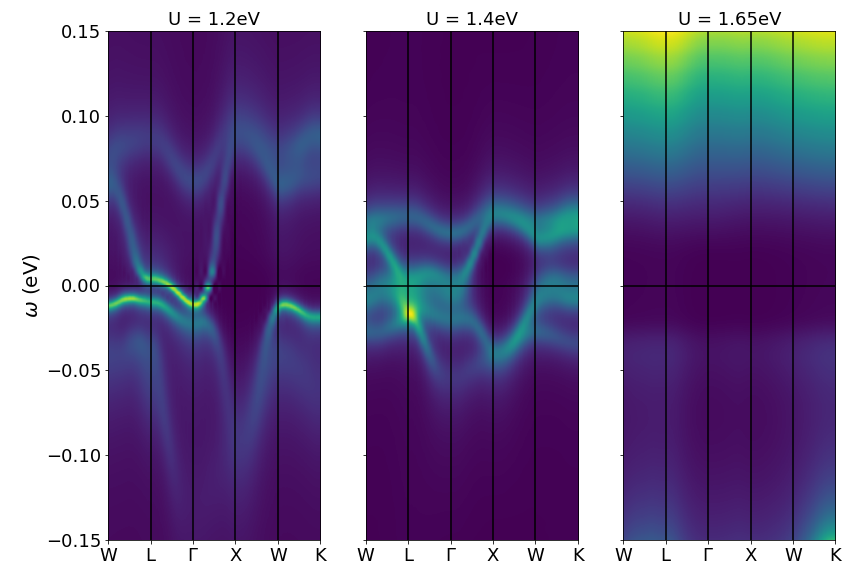}
		\caption{Momentum resolved spectral function at $U=1.2$\,eV, $U=1.4$\,eV, and $U=1.65$\,eV, calculated within sDMFT. Note that these calculations are in the magnetically ordered phase. Upon increasing $U$, bands get renormalized with a coherent quasi-particle up to 
        $U=1.6$\,eV. At $U=1.65$\,eV the Mott gap has fully opened, see right panel.}
		\label{fig:DMFT_bandstructure}
	\end{center}
\end{figure}

When a system is in a WSM phase, it shows a band crossing at the Fermi level for a specific k-point, while the bands are fully separated at all other k-points. At a band crossing like that, the electrons, the Weyl fermions of our system, have a high mobility. Looking at the momentum-resolved spectral function obtained from sDMFT, we can only see a trivial Mott-transition, see Fig.~\ref{fig:DMFT_bandstructure}. The bandstructures are obtained by analytically continuing the momentum dependent self energies
using TRIQS/MaxEnt~\cite{maxent}. 
First, the bands get renormalized and form a quasi-particle peak. Once the system reaches a critical interaction value, in the case of Y$_2$Ir$_2$O$_7$ at $U=1.6$\,eV, the system transitions from a metal to an insulator and the now incoherent Hubbard bands are pushed apart quickly as function of $U$. This result is in accordance with other findings in literature.

\begin{figure}
	\begin{center}
		\includegraphics[width=0.9\columnwidth]{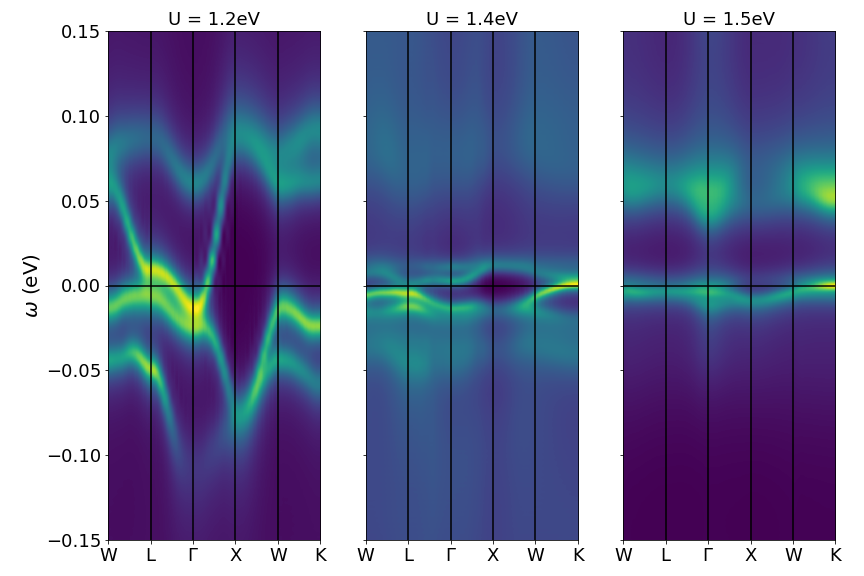}
		\caption{Momentum resolved spectral function at $U=1.2$\,eV, $U=1.4$\,eV, and $U=1.5$\,eV, calculated with the TRILEX approximation. The bandstructure for $U=1.2$\,eV is similar to the sDMFT bandstructure at the same interaction. At $U=1.4$\,eV the difference to sDMFT is very significant. }
		\label{fig:TRILEX_bandstructure}
	\end{center}
\end{figure}

\begin{figure}
	\begin{center}
		\includegraphics[width=0.8\columnwidth]{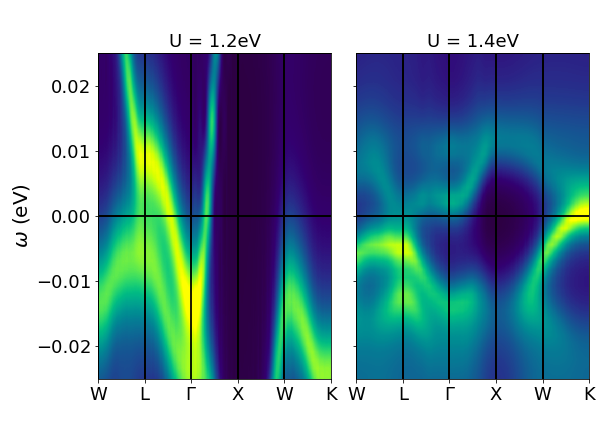}
		\caption{Zoomed-in TRILEX bandstructure for the interaction values $U=1.2$\,eV (left panel) and $U=1.4$\,eV (right panel) to better display the Weyl bandcrossing at the L high-symmetry point. }
		\label{fig:TRILEX_bandstructure_zoom}
	\end{center}
\end{figure}

The TRILEX solutions shown in Fig.~\ref{fig:TRILEX_bandstructure} behave very similar to sDMFT calculations up to intermediate interactions.
For probing higher interaction values, near the phase transition, the computational effort becomes exceedingly high. Therefore,  we are forced to rely on one-shot calculations in order to obtain insight in the physics near the phase transition. When inspecting the momentum-resolved spectral function, we see 
the expected band crossing at the L high-symmetry point. Especially for $U=1.4$\,eV, the characteristic band crossing with massless fermions is appearing. In order to see this better, we show in Fig.~\ref{fig:TRILEX_bandstructure_zoom} a zoom-in to low energies around the Fermi level.  However, the bands seem to also touch near the K high-symmetry point. This additional closing for the K high symmetry point could arise due to three different reasons. 

First, the reason could be the low resolution of the analytic continuation in combination with statistical noise in the QMC data
and therefore the bands only look like they are touching, when in reality there would be a gap. However, there is definitely a trend of gap narrowing happening at the K point, once the solution is allowed to show k-dependence. Even with a possible low resolution in the analytical continuation, it can not be neglected, that the bands are narrowing at this high symmetry point.
Another possibility is the fact that at this interaction regime we can only perform one-shot calculations and these do only display a trend of TRILEX, but not the physics of a fully converged TRILEX calculation. This is, however, difficult to rule out since full self-consistent TRILEX solutions at the phase transition point seem to be out of reach at the moment. 
The final explanation is that this is not an artefact but the result one actually gets from TRILEX, and calculating the triply irreducible vertex indeed leads to the band crossing at the L point as well as a band touching at the K point, prohibiting a WSM phase and only giving rise to a Weyl Metal phase. 
Regardless of the physics happening at the K high symmetry point, there is definitely a Weyl band crossing to be seen at the L high symmetry point and, hence, pyrochlore iridates have a Weyl Semimetal or at least a Weyl metal phase.
Further investigation at a larger interaction value $U=1.5$\,eV shows that the phase transition is likely to occur at a somewhat lower critical value as compared to the sDMFT calculations. Although we cannot calculate directly in the vicinity of the transition, by comparisons of sDMFT and TRILEX results for different interaction values $U$ we estimate the critical transition in TRILEX to be around $U\approx1.55$\,eV.

\section{Conclusions}

In this work we first investigated to the $j_\textrm{eff} = \frac{1}{2}$ model, and its qualitative equivalence to the three-orbital model including all $t_{2g}$ orbitals. This reduction to a single-orbital model is necessary in order to make TRILEX calculations computationally manageable. In the calculations of both single site DMFT as well as TRILEX the magnetic ordering seen in other works could be reproduced, and the transition to the antiferromagnetic all-in-all-out ordering confirmed. This symmetry breaking magnetic ordering is an essential requirement to topological properties in materials. 

With the sDMFT solutions not giving rise the a topological solution, the k-dependence of the self energy becomes a very important property to investigate in order to determine the Weyl phase. We could show that the self energy shows  increasing k-dependence for increasing interaction values. In particular, near the critical interaction value the k-dependence is very strong. This eventually can give rise to 
a Weyl-semimetal phase. 
It also aligns with the literature results of cluster DMFT where the solutions are k-dependent as well. 

The bandstructures of Y$_2$Ir$_2$O$_7$ show the typical bandcrossing occuring due to Weyl cones. With the single-shot TRILEX calculations the WSM cannot be confirmed to highest certainty. This uncertainty is due to the band narrowing near the K high-symmetry points. Regardless of that, we could show the presence of Weyl band crossings. Therefore, the pyrochlore iridates host a Weyl semimetal or at least a Weyl metal phase.

\acknowledgments
We thank Thomas Sch\"afer and Nils Wentzell for insightful discussions on TRILEX. This work has been supported by the Austrian Science Fund (FWF), grant Y746. A part of the calculations have been performed on the Vienna Scientific Cluster (VSC). 

\appendix

\bibliography{Y2Ir2O7}

\end{document}